\documentclass[preprint,12pt]{elsarticle}

\usepackage{amssymb}
\usepackage{graphicx}
\usepackage{amsmath}
\DeclareGraphicsExtensions{.pdf,.jpeg,.png}
\usepackage{array}
\usepackage{mathrsfs}
\usepackage{bm}
\usepackage{color}
\usepackage{hyperref}
\usepackage{lineno}

\def\##1{\bm {#1}}
\def\=#1{\underline{\underline #1}}
\def\~#1{\tilde{#1}}
\def\^#1{\breve{#1}}
\def\`#1{{#1^\prime}}
\def\:#1{#1^{\prime\prime}}
\def\h#1{\hat{\bm{#1}}}
\def\s#1{\bm{\mathcal{#1}}}
\def\.{\mbox{ \tiny{$^\bullet$} }}

\journal{Physics Letters A}

\begin{document}

\begin{frontmatter}
\title{Scattering characteristics of relativistically moving concentrically layered spheres}

 \author[label1]{Timothy J. Garner}
\author[label2]{Akhlesh Lakhtakia}
\author[label1]{James K. Breakall}
\author[label3]{Craig F. Bohren}
 \address[label1]{Department of Electrical Engineering, Pennsylvania State University, University Park, PA 16802, USA, tjg236@psu.edu}
 \address[label2]{Department of Engineering Science and Mechanics, Pennsylvania State University, University Park, PA 16802, USA}
 \address[label3]{Department of Meteorology, Pennsylvania State University, University Park, PA 16802, USA}

\begin{abstract}
The energy extinction cross section of a concentrically layered sphere varies with velocity as the Doppler shift moves the spectral content of the incident signal  in the sphere's co-moving inertial reference frame toward or away from resonances of the sphere.  Computations for hollow gold nanospheres show that the energy extinction cross section is high when the Doppler shift moves the incident signal's spectral content in the co-moving frame near the wavelength of the sphere's localized surface plasmon resonance.  The energy extinction cross section of a three-layer sphere consisting of an olivine-silicate core surrounded by a porous and a magnetite layer, which is used to explain extinction caused by interstellar dust, also depends strongly on velocity.   For this sphere,  computations show that the energy extinction cross section is high when the Doppler shift moves the spectral content of the incident signal near either of  olivine-silicate's two localized surface phonon resonances at 9.7~$\mu$m and 18~$\mu$m.  

\end{abstract}


\end{frontmatter}


\section{Introduction}
Researchers have investigated planewave scattering by  electrically small objects in uniform translatory motion at relativistic speeds. Typically, the investigations are done with a frame-hopping technique in which two inertial reference frames, one affixed to the laboratory and the other to the moving object, are used \cite{Shiozawa1968,Lakhtakia1991a, Lakhtakia1991b}.  The frequency-domain constitutive relations of the material of the object as well as the object's dimensions are specified in the co-moving frame. The electrically small object is modeled as an assembly of two co-located  dipoles, one electric and the other magnetic. The Lorentz transformation is used to transform the incident and the scattered field from one reference frame to the other. 

An object may be  electrically small only if the wavelength of the incident plane wave in the co-moving frame exceeds a threshold value \cite{vDH}. At shorter
wavelengths, the same object may even be electrically large. An object that is electrically small in the laboratory frame when at rest may be electrically large in its co-moving frame when it is in motion due to the Doppler effect.  Furthermore, with modern instrumentation, spectroscopic information about an object is usually obtained by illuminating it with a pulse rather than a plane wave.  A pulse has finite duration and can have a large bandwidth, unlike a plane wave which is eternal and monochromatic. Therefore, the time-domain scattering response of any object, moving or not, is desirable.

In this Letter, we use the frame-hopping technique to investigate the effects of uniform translational motion on the energy extinction, energy absorption, and total energy scattering cross sections \cite{Garner2017_3} of inhomogenous objects.  We chose concentrically layered spheres as suitable examples because an analytical solution for planewave scattering by these objects exists \cite{AK,Bhandari} in the co-moving frame, thereby avoiding inaccuracies arising from the implementation of numerical techniques such as the finite-difference time-domain method \cite{KL,TH}.    We examined the energy cross sections of a hollow gold sphere as  functions of velocity and inner diameter, the outer diameter being fixed. We also computed the energy cross sections as functions of velocity of a three-layer sphere proposed by Voshchinnikov \textit{et al.} \cite{Voshchinnikov2017} to explain observations of infrared extinction due to interstellar dust.  Energy cross sections are appropriate for pulse scattering just as power cross sections \cite[Sec.~3.4]{Bohren} are for planewave scattering.
   
\section{Method}

We used the frame-hopping technique \cite{Garner2017_1} to compute the electric and magnetic fields of the  scattered signals arising from the illumination of a uniformly translating concentrically layered sphere by a signal of finite duration.  The techinque has four steps:  (i) define the electric and magnetic fields of the incident signal in the laboratory inertial reference frame $K'$, (ii) use the Lorentz transformation to express the fields of the incident signal in the   co-moving inertial reference frame $K$, (iii) compute the electric and magnetic fields of the scattered signals in $K$, and (iv) transform the fields of the scattered signals to $K'$ with the inverse of the Lorentz transformation used in Step~(ii).  

The incident signal was chosen to be a  plane wave with its amplitude modulated by a Gaussian pulse.  The signal travels in the $+z'$ direction with its electric field  in the $x'$ direction.  The electric and magnetic fields of the incident signal in $K'$ are
\begin{subequations}
\begin{align}
	\s E'_{inc} \left(\bm{r}',t'\right)&= \h x' \cos(2\pi \nu'_c\tau') \exp\left(-\frac{\tau'^2 }{ 2\sigma'^2}\right),   \\
	\s B'_{inc} \left(\bm{r}',t'\right)&=\frac{\#{\hat{z}}'\times \s E'_{inc} \left(\#r',t'\right)}{c},
\end{align}
\end{subequations}
where $\#r'$ is the position vector, $t'$ is time,
$\tau' = t' - (\h z'\cdot \#r')/c $, $\sigma'$ is the width parameter of the Gaussian pulse, $\nu'_c$ is the carrier frequency,  $c$ is the speed of light in free space, and every unit vector is decorated by a caret.

We converted the electromagnetic fields of the incident signal to the frequency domain by using a discrete Fourier transform and computed the scattered field phasors using an analytical technique \cite{AK,Bhandari} based on the Lorenz--Mie series solution \cite{Bohren} in Step~(ii).  Details of the transformations between time and frequency domains are available elsewhere \cite{Garner2017_3,Garner2017_1}.  As with a solid, homogenous sphere, the power  scattering, power absorption, and power extinction cross sections in $K$ were computed using the coefficients of the scattered field phasors in the Lorenz--Mie solution \cite[Eqs.~(4.61, 62)]{Bohren}.  To find these coefficients, we used  (i) expressions from Bohren and Huffman \cite[Eq.~(8.2)]{Bohren} for hollow gold spheres and  (ii) expressions from Shore \cite{Shore2015} for three-layer spheres.  The energy absorption and energy extinction cross sections were determined from the power absorption and power extinction cross sections \cite[Eqs.~(34),~(40)]{Garner2017_3}, respectively.  We computed the total energy scattering cross section by determining the scattered signals in all scattering directions and numerically integrating the scattered power with respect to time to get the scattered energy density (with units of energy per solid angle) in all directions.  The total scattered energy was calculated by numerically integrating the scattered energy density using 41-point Gauss-Kronrod quadrature \cite{Piessens1983}\cite[pp.~153-155]{Kahaner1989} over $\theta'$ and 32-point rectangular integration over $\phi'$ \cite{Kahaner1989}, as described elsewhere  \cite[Sec.~IIA]{Garner2017_3}

\begin{sloppypar}
We used measured, wavelength-dependent constitutive parameters for bulk gold \cite{Hagemann1975},  olivine silicate (MgFeSiO$_4$) \cite{Dorschner1995}, and magnetite (Fe$_3$O$_4$) \cite{Triaud}.   The parameters for gold were obtained from the online database at \url{http://www.refractiveindex.info} and those for olivine silicate and magnetite from the University of Jena website, \url{http://www.astro.uni-jena.de/Laboratory/OCDB/index.html}.
\end{sloppypar}

The \textit{normalized} total energy scattering, absorption, and extinction cross sections in $K'$ are defined as \cite{Garner2017_3}
\begin{subequations}
\begin{align}
Q'_{sca} = \frac{W'_{sca}}{U'_{inc}A},\\
Q'_{abs} = \frac{W'_{abs}}{U'_{inc}A},\\
Q'_{ext} = \frac{W'_{ext}}{U'_{inc}A},
\end{align}
\end{subequations}
where $U'_{inc}$ is the energy density of the incident signal with units of energy per area; $A$ is the physical cross-sectional area of the sphere in $K'$; and $W'_{sca}$, $W'_{abs}$, and $W'_{ext}$ are the total energies scattered by the object, absorbed by the object, and removed from the incident signal by the object, respectively.    Analagous normalized energy cross sections may also be defined in $K$.  

We also define the standard normalized power extinction cross section in $K$ as  \cite{Garner2017_3}
\begin{equation}
\~Q_{ext}(\lambda) = \frac{\~P_{ext}(\lambda)}{\~U_{inc}(\lambda)A},
\end{equation}
where $\~P_{ext}(\lambda)$ is the time-averaged power removed from a plane wave of free-space wavelength $\lambda$ and $\~U_{inc}(\lambda)$ is the power density of the incident plane wave.

\section{Numerical Results}
\subsection{Hollow gold nanophere}
\begin{figure}
   \centering
   \includegraphics[width=5.5in]{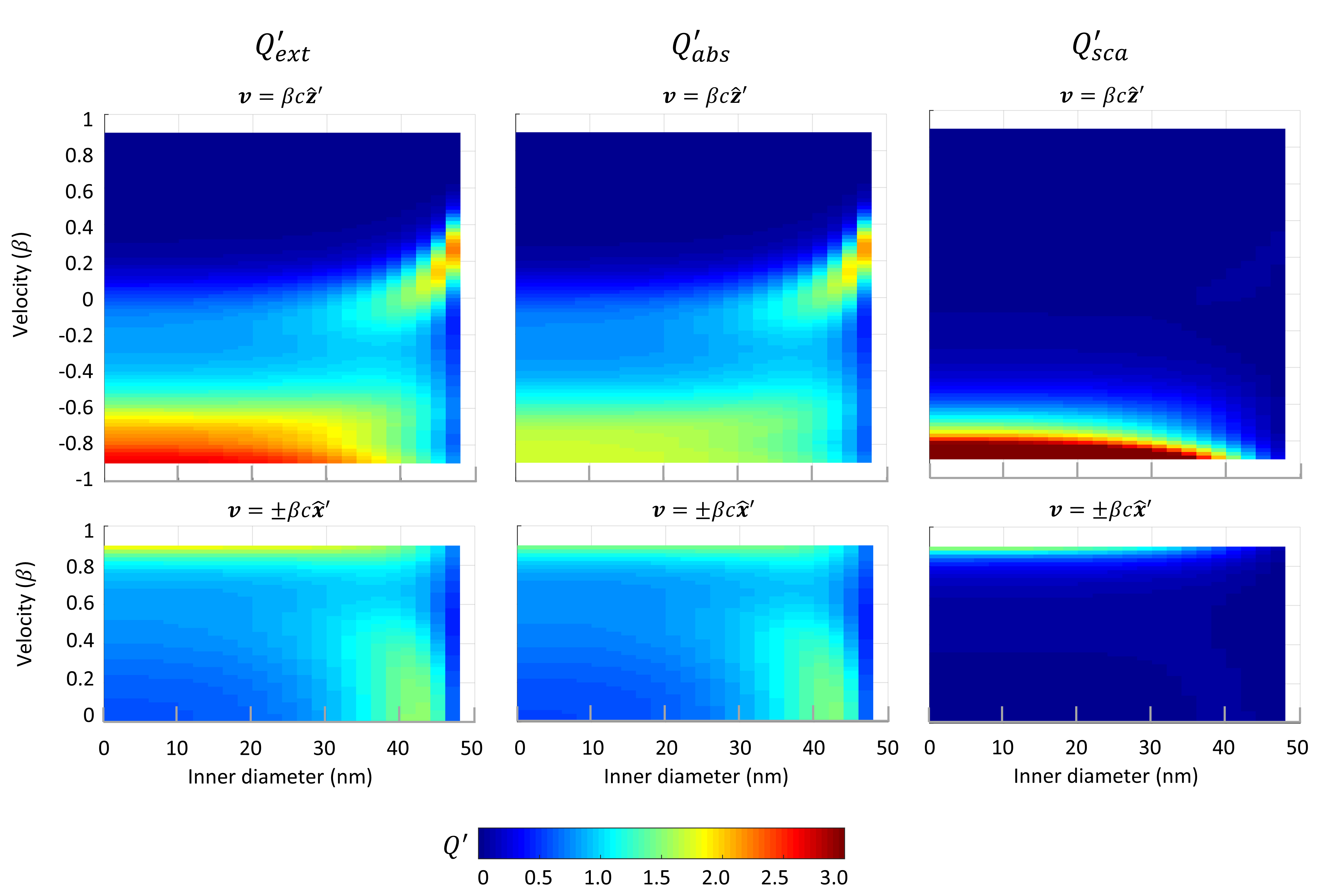}
      \caption{Normalized energy extinction ($Q'_{ext}$), energy absorption ($Q'_{abs}$), and total energy scattering ($Q'_{sca}$) cross sections in $K'$ of a hollow gold sphere as functions of velocity $\beta{c}\h z'$ or $\pm\beta{c}\h x'$
      and the inner diameter in $K$, when the outer diameter in $K$ is fixed at 50~nm.  The carrier wavelength of the incident signal in $K'$ is $\lambda'_c=550$~nm. The permittivity of bulk gold was used in the computations.   
              }
         \label{FigQext_p_gold}
   \end{figure}

Fig. \ref{FigQext_p_gold} shows the normalized energy extinction, energy absorption, and total energy scattering cross sections of a hollow gold sphere as  functions of velocity and inner diameter (in $K$), when the outer diameter in $K$ is fixed at 50~nm. In $K'$, the carrier wavelength of the incident signal is $\lambda'_c=550$~nm, and the width parameter of the Gaussian function is $\sigma'=1.833$~fs. 

\begin{figure}
   \centering
   \includegraphics[width=3.5in]{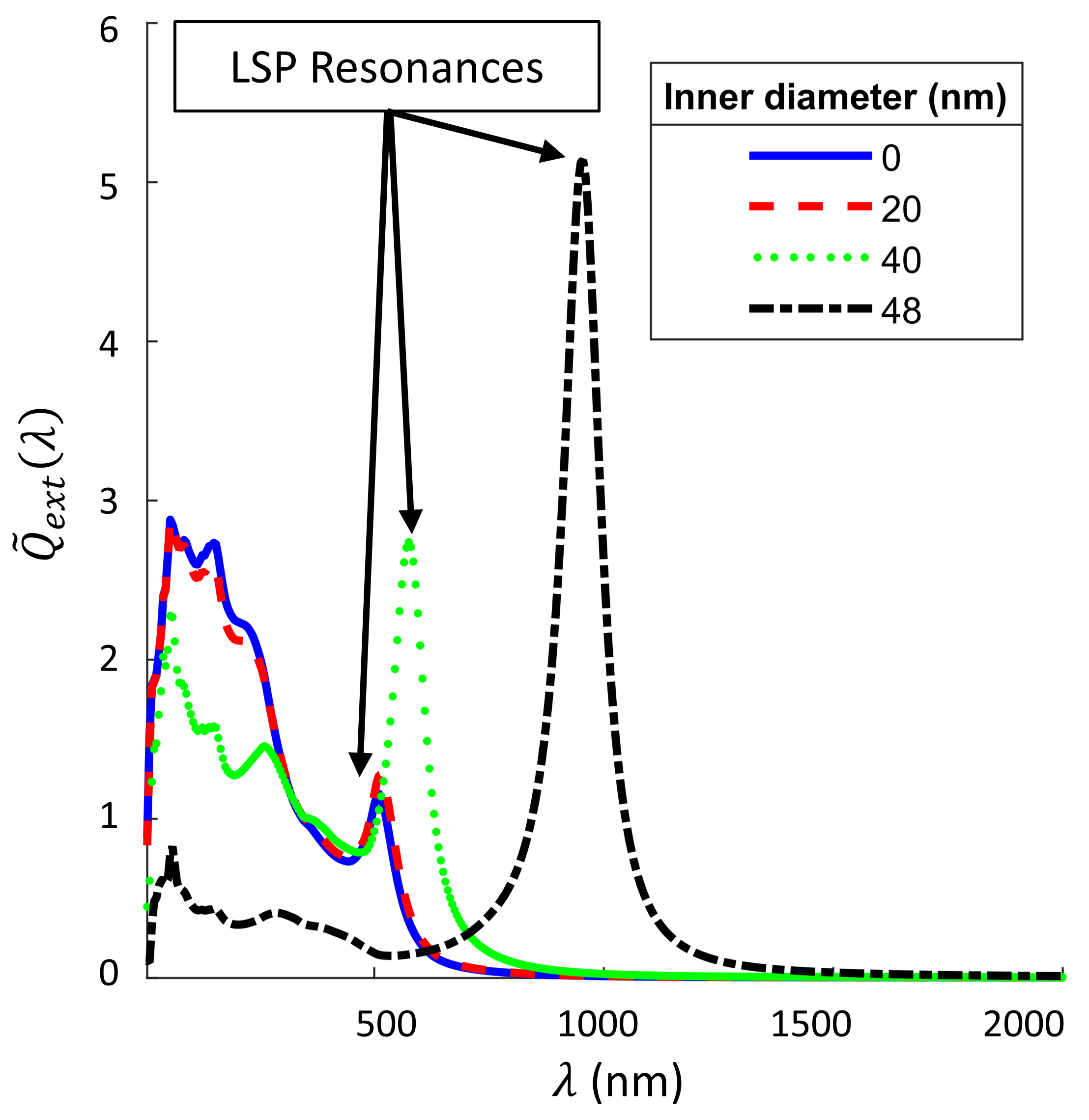}
      \caption{Normalized power extinction cross section $\~Q_{ext}(\lambda)$  vs. wavelength $\lambda$ in $K$ of a hollow gold sphere for various inner diameters, when the outer diameter is fixed at 50~nm.  The permittivity of bulk gold was used in the computations.  The locations of the localized surface plasmon (LSP) resonances are indicated.  
              }
         \label{FigAu_Qextw}
   \end{figure}

All three cross sections, $Q'_{ext}$, $Q'_{abs}$, and $Q'_{sca}$, are high when the sphere advances toward the source of the incident signal at speeds approaching $c$.  This occurs because the Doppler shift increases the electrical size of the sphere in $K$ as the sphere advances toward the source at speeds approaching $c$.   $Q'_{ext}$, $Q'_{abs}$, and $Q'_{sca}$ are near zero when the sphere recedes from the source at speeds approaching $c$, because the electrical size of the sphere in $K$ goes to zero. $Q'_{ext}$ and $Q'_{abs}$ also have high values in the region where the inner diameter ranges from about 34--48~nm and the sphere recedes from the source of the incident signal at velocities ranging from 0 to $0.3c\h z'$.  The cause of this high-$Q'_{ext}$ regime may be understood by considering the  normalized power extinction cross section $\~Q_{ext}(\lambda)$ in $K$, as shown in Fig. \ref{FigAu_Qextw}.  As the inner diameter increases, the strength of the localized surface plasmon resonance \cite[Sec.~2.4.1]{Kreibig1995} increases and its wavelength shifts from about $515$~nm to about $955$~nm.  $Q'_{ext}$ has a maximum when the Doppler shift causes the spectral content of the incident signal in $K$ to coincide with the localized surface plasmon resonance.  

Bulk permittivity becomes inapplicable to a metal shell whose thickness $h$ is less than the mean free path of  electrons in the bulk metal.  Therefore, we also computed the normalized energy cross sections of a hollow gold sphere using a modified permittivity function for small spheres given by Kreibig and Fragstein \cite{Kreibig1969}. In a particle whose linear dimensions are on the order of the mean free path, free electrons may collide with the edges as well as the lattice defects in the material \cite[pp. 78--85]{Kreibig1995}. The collisions with the boundary decrease the mean free path and increase the collision frequency in the Drude  model of the permittivity.

We assumed that the frequency-dependent relative permittivity of bulk gold given by Hagemann \textit{et al.} \cite{Hagemann1975} is a combination of a Drude component and a bound-electron component,
\begin{equation}
\epsilon_{bulk}(\nu) = \epsilon_{bound}(\nu) + \epsilon_{Drude, bulk}(\nu),
\end{equation}
and that the Drude component of this relative permittivity is \cite[Eq.~2]{Ordal1985}
\begin{equation}
\epsilon_{Drude, bulk}(\nu) = \epsilon_\infty - \frac{\nu_p^2}{\nu^2 + i\nu\nu_{\tau, bulk}}
\end{equation}
with plasma frequency $\nu_p= 2.17\times 10^{15}$~s$^{-1}$ and collision frequency $\nu_{\tau, bulk}=6.48\times 10^{12}$~s$^{-1}$ \cite[Table~1]{Ordal1985}.   In a solid, homogenous sphere of radius $R$, the collision  frequency becomes \cite[Eq.~(10)]{Kreibig1969}
\begin{equation}
\nu_{\tau, sphere} = \nu_{\tau, bulk} + \frac{v_F}{{2\pi}R},
\end{equation}
where for gold, the Fermi velocity $v_F=2\pi l_{\infty}\nu_{\tau,bulk}=1.709\times 10^6$~m/s as the mean free path $l_{\infty}=42$~nm \cite[Table~2.3]{Kreibig1995}.  We used the correction for the collision frequency
\begin{equation}
\label{eq:nushell}
\nu_{\tau, shell} = \nu_{\tau, bulk} + \frac{v_F}{{2\pi}h}.
\end{equation}We corrected the relative permittivity by replacing the Drude component for the bulk material with the Drude component for the shell 
\begin{equation}
\epsilon_{shell}(\nu) = \epsilon_{bound}(\nu) + \epsilon_{Drude,shell}(\nu),
\end{equation}
where 
\begin{equation}
\epsilon_{Drude, shell}(\nu) = \epsilon_\infty - \frac{\nu_p^2}{\nu^2 + i\nu\nu_{\tau, shell}}.
\end{equation}

\begin{figure}
   \centering
   \includegraphics{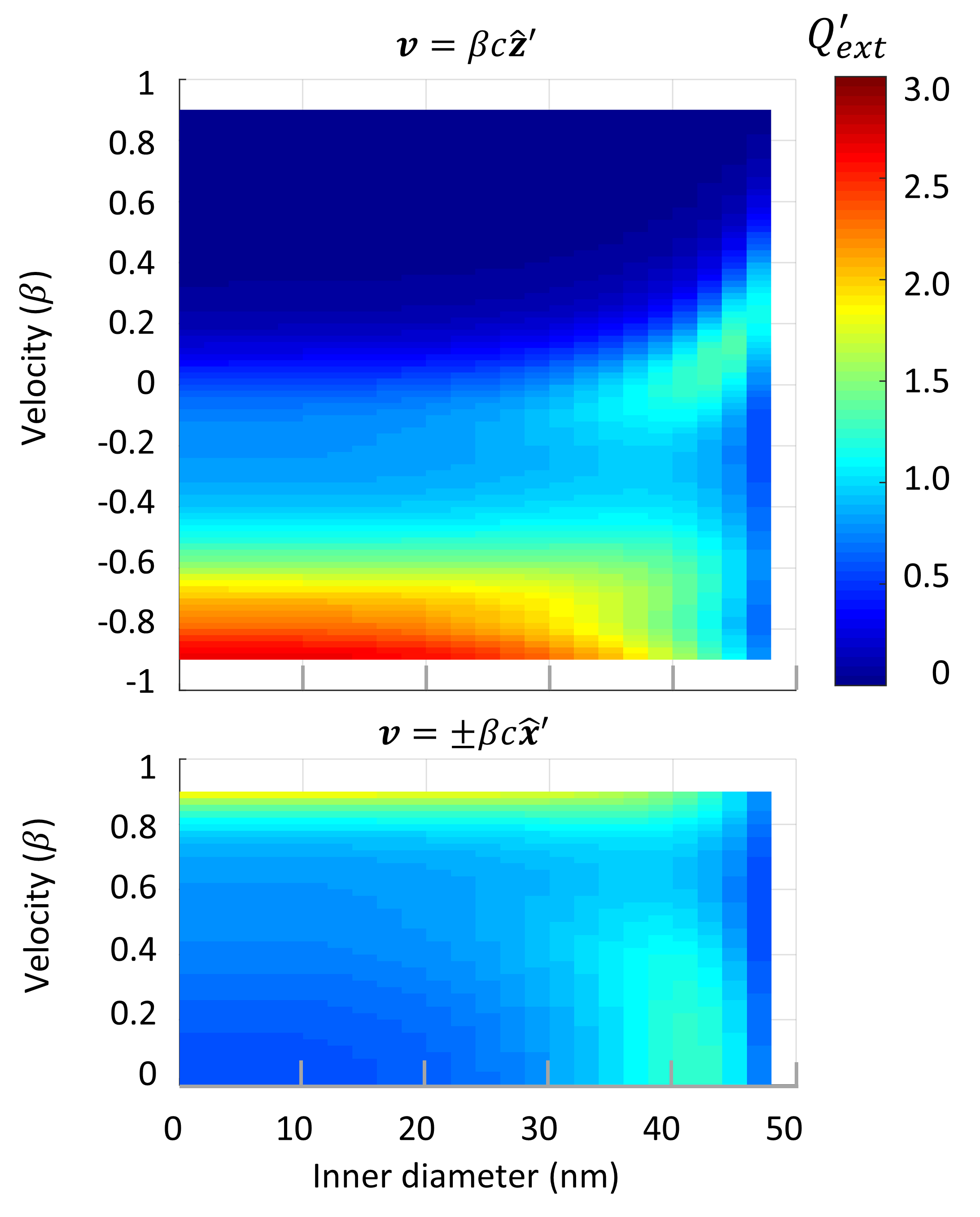}
         \caption{Normalized energy extinction cross section $Q'_{ext}$
         as a function of velocity $\beta{c}\h z'$ or $\pm\beta{c}\h x'$
      and the inner diameter in $K$, when the outer diameter in $K$ is fixed at 50~nm.  The carrier wavelength of the incident signal in $K'$ is $\lambda'_c=550$~nm. The permittivity of  gold was adjusted for the reduction in the  mean free path of electrons for these data.   
                            }
         \label{FigQext_p_gold_MFP}
   \end{figure}

Numerical results with the relative permittivity of gold adjusted for the reduction in mean free path  are shown in Fig. \ref{FigQext_p_gold_MFP}. The mean-free-path correction reduces the strength of the localized surface plasmon resonance, which leads to a reduction in $Q'_{ext}$ when the sphere directly recedes from the source and the inner diameter approaches the outer diameter.  

\subsection{Three-layer sphere}

We also computed  all three normalized energy cross sections of a three-layer sphere proposed by Voshchinnikov \textit{et al.} \cite{Voshchinnikov2017} to explain infrared extinction caused by interstellar dust.  The object consists of an olivine-silicate core and an outer layer of magnetite which is separated from the core by a porous layer that is approximated as free space.  The diameter of the core is 342~nm, and  both the porous and the magnetite layers are 3~nm thick.  

\begin{figure}
   \centering
   \includegraphics[width=3.5in]{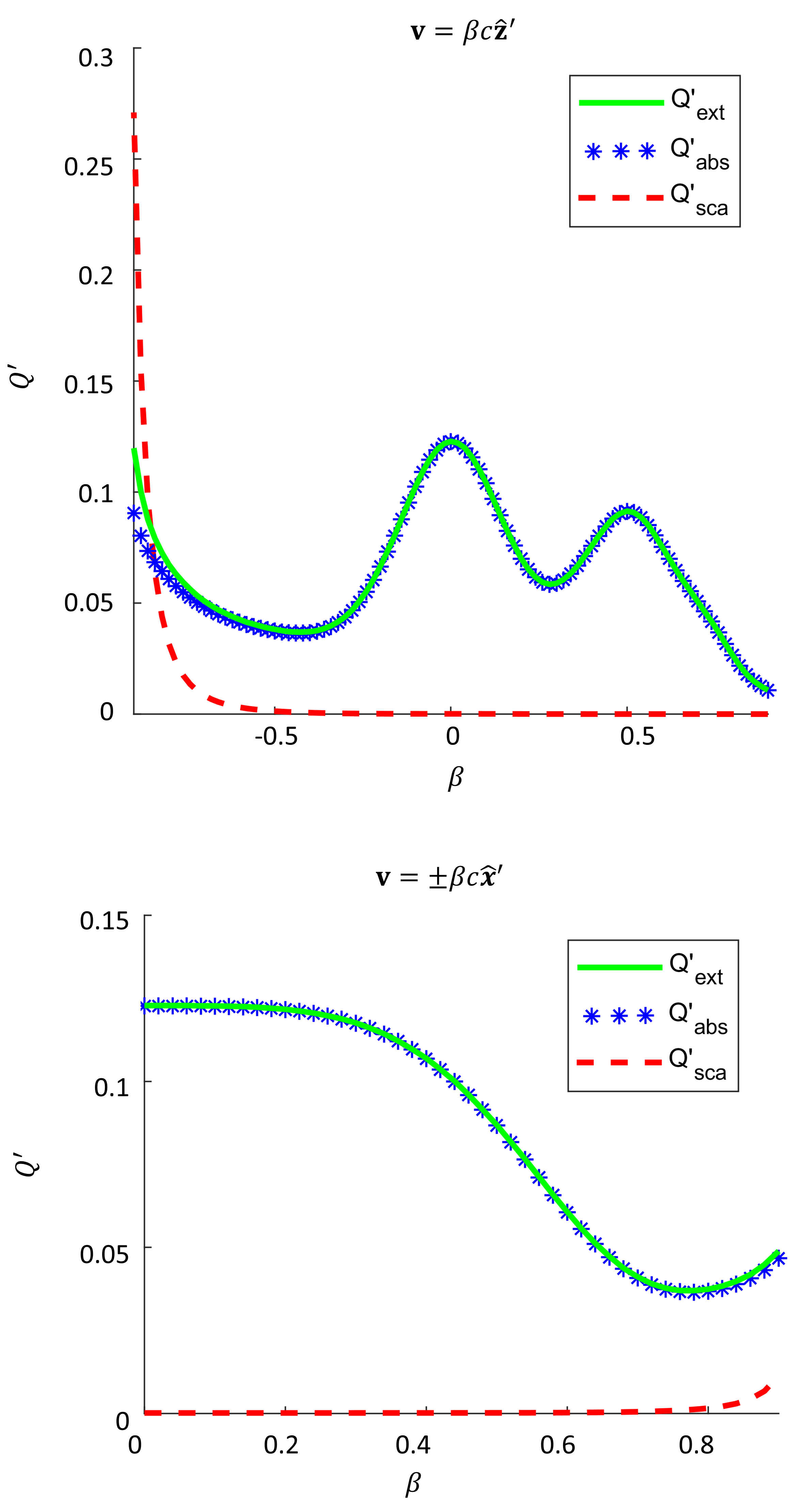}
      \caption{Normalized energy extinction ($Q'_{ext}$), energy absorption ($Q'_{abs}$), and total energy scattering ($Q'_{sca}$) cross sections of the three-layer sphere.}
         \label{Fig_Voshchinnikov_Q}
   \end{figure}

Fig. \ref{Fig_Voshchinnikov_Q} shows $Q'_{ext}$, $Q'_{abs}$, and $Q'_{sca}$ of this sphere as  functions of speed for the three-layer sphere  advancing directly toward and receding directly from the source of the incident signal (top panel) and moving transversely to the source of the incident signal (bottom panel).  The carrier wavelength of the signal in $K'$ is $\lambda'_c = 10$~$\mu$m, and the width parameter of the Gaussian function is $\sigma'=33.3$~fs.  In $K$, this sphere has localized surface phonon resonances \cite{Ruppin1970} at $\lambda=9.7$~$\mu$m and 18~$\mu$m \cite{Voshchinnikov2017}.  

When $\#v=\beta{c}\h z'$, $Q'_{ext}$ is close to a maximum when the object is stationary, because the spectral content of the incident signal in $K$ coincides with the silicate resonance at 9.7~$\mu$m.  As it recedes from the source with increasing velocity, $Q'_{ext}$ decreases before increasing again to another local maximum at $\#v = 0.5c\h z'$.  At this velocity, the carrier wavelength $\lambda_c$ in $K$ is shifted to about $18$~$\mu$m and overlaps with the silicate resonance at 18~$\mu$m \cite[Fig. 3]{Voshchinnikov2017}. All three cross sections decrease and approach zero when the sphere recedes from the source of the incident signal with speeds approaching $c$, because the electrical size of the sphere in $K$ goes to zero.  When the sphere advances directly toward the source of the incident signal, $Q'_{ext}$ first decreases with increasing speed, because the Doppler shift moves the incident signal in $K$ away from the silicate resonance at 9.7~$\mu$m.  When the speed of the advancing sphere increases beyond $\#v \approx -0.4c\h z'$, $Q'_{ext}$, $Q'_{abs}$, and $Q'_{sca}$ increase because the Doppler shift increases the electrical size of the sphere in $K$.  

When the sphere moves along the $x'$ axis, i.e., perpendicular to the incident signal's direction of propagation, $Q'_{ext}$  decreases as the speed increases before beginning to rise again at about $\#v = \pm 0.8c\h x'$.  This occurs because the transverse Doppler effect shifts the spectral content of the incident signal to shorter wavelengths in $K$, away from the 9.7-$\mu$m localized surface phonon resonance.  $Q'_{ext}$, $Q'_{abs}$, and $Q'_{sca}$ begin to increase when the sphere moves in the $\pm x'$ direction with $\#v=\pm 0.8c\h x'$.  

\section{Concluding Remarks}
Uniform translation at relativistic speeds affects the energy extinction, and energy absorption cross sections in $K'$ of an object by shifting the spectral content of the incident signal in $K$  either toward or away from resonances observed when the object is stationary.  This is evident in the computed results for the energy extinction cross section $Q'_{ext}$ of a hollow gold sphere when the diameter of the void is 34--48~nm.  The energy extinction cross section is greatest for these   spheres when they recede from the source at such a velocity that the spectral content of the incident signal is shifted to the localized surface plasmon resonance in $K$.  The velocity at which the maximum occurs varies with the inner diameter of the sphere, because the frequency of the localized surface plasmon resonance is a function of the inner diameter.  

This effect is also seen in the normalized energy extinction cross section of the three-layer dust particle model.  In $K$, olivine silicate has localized surface phonon resonances at 9.7~$\mu$m and 18~$\mu$m.  When the particle is at rest, the spectrum of incident signal is centered at $\lambda_c=10$~$\mu$m, close to the resonance at 9.7~$\mu$m in $K$.  As the particle recedes from the source with increasing speed, $Q'_{ext}$ first decreases before rising to another peak when $\#v = 0.5c\h z'$.  This second peak occurs when the relativistic Doppler shift moves spectral content of the incident signal in $K$ to the resonance at 18~$\mu$m.  

For both objects examined, all three energy cross sections go to zero as they recede from the source of the incident signal with speeds approaching $c$.  This happens because the electrical size of the object goes to zero in $K$.  As the object advances directly toward the source of the incident signal at speeds approaching $c$, the cross sections increase because the electrical size of the particle increases in $K$.

\section*{Acknowledgments}
The authors thank Prof. Thomas Henning (Max-Planck-Institut f\"ur Astronomie, Heidelberg) for information of interstellar dust.
T. J. G. was supported by a Graduate Excellence Fellowship from the College of Engineering,
Pennsylvania State University. A. L. is grateful for the support of the Charles
Godfrey Binder Endowment at the Pennsylvania State University.





\section*{References}
\begin{thebibliography}{99}
\expandafter\ifx\csname url\endcsname\relax
  \def\url#1{\texttt{#1}}\fi
\expandafter\ifx\csname urlprefix\endcsname\relax\def\urlprefix{URL }\fi
\expandafter\ifx\csname href\endcsname\relax
  \def\href#1#2{#2} \def\path#1{#1}\fi

\bibitem{Shiozawa1968}
T.~Shiozawa, Electromagnetic scattering by a moving small particle, Journal of
  Applied Physics 39 (1968) 2993--2997.

\bibitem{Lakhtakia1991a}
A.~Lakhtakia, V.~V. Varadan, V.~K. Varadan, Plane wave scattering response of a
  simply moving, electrically small, chiral sphere, Journal of Modern Optics 38
  (1991) 1033--1036.

\bibitem{Lakhtakia1991b}
A.~Lakhtakia, Dyadic procedure for planewave scattering by simply moving,
  electrically small, bianisotropic spheres, Journal of Modern Optics 38 (1991)
  1033--1036.
  
\bibitem{vDH}
H. C.  van de Hulst, Light Scattering by Small Particles, Sec. 6.4, Dover Press, 1981.

\bibitem{Garner2017_3}
T.~J. Garner, A.~Lakhtakia, J.~K. Breakall, C.~F. Bohren, Lorentz invariance of
  absorption and extinction cross sections of a uniformly moving object, arXiv:1710.03859.

\bibitem{AK}
A. L. Aden, M. Kerker, Scattering of electromagnetic waves from two concentric spheres, Journal of Applied Physics 22 (1951) 1242--1246.

 
\bibitem{Bhandari}
R.~Bhandari, Scattering coefficients for a multilayered sphere: analytical expressions and algorithms, Applied Optics 24  (1985) 1960--1967.
  
  \bibitem{KL}
K. S. Kunz and R. J. Luebbers, The Finite Difference Time Domain
Method for Electromagnetics, CRC Press, Boca Raton, 1993.

\bibitem{TH}
A. Taflove and S. C. Hagness, Computational Electrodynamics: The
Finite-Difference Time-Domain Method, 3rd ed., Artech House, Boston,
2005.


\bibitem{Voshchinnikov2017}
N.~V. Voshchinnikov, T.~Henning, V.~B. Il'in, Mid-infrared extinction and fresh
  silicate dust toward the galactic center,  Astrophysical Journal 837 (2017) 25.

\bibitem{Bohren}
C.~F. Bohren, D.~R. Huffman, Absorption and Scattering of Light by Small
  Particles, Wiley, New York, 1983.

\bibitem{Garner2017_1}
T.~J. Garner, A.~Lakhtakia, J.~K. Breakall, C.~F. Bohren, Time-domain
  electromagnetic scattering by a uniformly translating sphere, Journal of the
  Optical Society of America A 34 (2017) 270--279.

\bibitem{Shore2015}
R.~A. Shore, Scattering of an electromagnetic linearly polarized plane wave by
  a multilayered sphere, Antennas and Propagation Magazine (2015) 69--116.

\bibitem{Piessens1983}
R.~Piessens, E.~{de Doncker-Kapenga}, C.~W. {\"U}berhuber, D.~K. Kahaner,
  QUADPACK: A Subroutine Package for Automatic Integration, Springer, Berlin,
  1983.

\bibitem{Kahaner1989}
D.~Kahaner, C.~Moler, S.~Nash, Numerical Methods and Software, Prentice Hall,
  Englewood Cliffs, New Jersey, 1989.

\bibitem{Hagemann1975}
H.~J. Hagemann, W.~Gudat, C.~Kunz, Optical constants from the far infrared to
  the x-ray region: Mg, Al, Cu, Ag, Au, Bi, C, and Al$_2$O$_3$, Journal of the
  Optical Society of America 65 (1975) 742--745.

\bibitem{Dorschner1995}
J.~Dorschner, B.~Begemann, T.~Henning, C.~J{\"a}ger, H.~Mutschke, Steps toward
  interstellar silicate mineralogy II. Study of Mg-Fe-silicate glasses of
  variable composition, Astronomy and Astrophysics 300 (1995) 503--520.

\bibitem{Triaud}
A.~H. Triaud, n-k data of iron oxides,
\url{http://www.astro.uni-jena.de/Laboratory/OCDB/mgfeoxides.html}.

\bibitem{Kreibig1995}
U.~Kreibig, M.~Vollmer, Optical Properties of Metal Clusters, Springer, Berlin,
  1995.

\bibitem{Kreibig1969}
U.~Kreibig, C.~v.~Fragstein, The limitation of electron mean free path in small
  silver particles, Zeitschrift f\"ur Physik 224 (1969) 307--323.



\bibitem{Ordal1985}
M.~A. Ordal, R.~J. Bell, R.~W. {Alexander, Jr.}, L.~L. Long, M.~R. Querry,
  Optical properties of fourteen metals in the infrared and far infrared: Al,
  Co, Cu, Au, Fe, Pb, Mo, Ni, Pd, Pt, Ag, Ti, V, and W, Applied Optics 24
  (1985) 4493--4499.

 \bibitem{Ruppin1970}
R. Ruppin and R. Englman, Optical phonons of small crystals, Reports on Progress in Physics 33 (1970) 149--196.
\end{thebibliography}
\end{document}